\documentclass[aps,prl,showpacs,superscriptaddress,twocolumn]{revtex4-1}
\usepackage[english]{babel}
\usepackage{graphicx}
\usepackage{stmaryrd}
\usepackage{amssymb}
\usepackage{amsfonts}
\usepackage{amsmath}
\usepackage{color}
\usepackage{xspace}

\usepackage{bm}
\usepackage{color}
\definecolor{lightblue}{rgb}{0.13, 0.26, 0.99}

\usepackage{hyperref}

\newcommand{\dw}{d_{\mathrm{w}}}

\begin{document}

\title{Negative Coulomb Drag in Coupled Quantum Wires} 
\author{Shunsuke C. Furuya}
\affiliation
{Department of Quantum Matter Physics, University of Geneva,24 Quai
Ernest-Ansermet 1211 Geneva, Switzerland}
\author{Hiroyasu Matsuura}
\affiliation{Department of Physics, University of Tokyo, 7-3-1 Hongo, Bunkyo-ku, Tokyo 113-0033, Japan}
\author{Masao Ogata}
\affiliation{Department of Physics, University of Tokyo, 7-3-1 Hongo, Bunkyo-ku, Tokyo 113-0033, Japan}
\date{\today}
\begin{abstract}
 We present a theory of negative Coulomb
 drag in capacitively coupled quantum wires based on the
 commensurability of the electron density and the long-range nature
 of the Coulomb interaction in the Tomonaga-Luttinger liquid.
 The commensurability introduces a notion of doped particles and holes.
 We point out that the long-range interaction allows a particle-hole
 pairing over the wires and that the particle-hole pairing brings about
 the positive drag of holes, that is, the negative drag.
\end{abstract}
\maketitle

Understanding many-body interaction effects is a longstanding objective of condensed matter physics as typified by the Fermi liquid theory~\cite{AGD}. 
The Fermi liquid theory however fails in strongly correlated systems, especially in one-dimensional (1D) systems~\cite{Giamarchi_book}.
Since any particle in 1D space cannot overtake others in front of it,
the restricted 1D geometry provokes characteristic transport phenomena
such as the quantization of conductance~\cite{Tarucha_G,Ogata_dirty_TLL} and 
diffusive transports compatible with integrability~\cite{Sirker_xxz,Hild_transport}.
In 1D systems, instead of the Fermi liquid,
the collective motion of the particle-hole pair in the so-called Tomonaga-Luttinger liquid (TLL)
is the most elementary excitation~\cite{Giamarchi_book, Haldane_bosonization}.

The TLL has a great advantage of incorporating interactions into a single parameter $K$ known as the TLL parameter. 
It is non-interacting for $K=1$, repulsive for $K<1$ and attractive for $K>1$~\cite{Giamarchi_book}. 
Furthermore $K$ is controllable with external parameters.
In the field of quantum magnetism, the experimental controllability of $K$ with the magnetic field is fully exploited
to simulate itinerant boson systems~\cite{Giamarchi_TLL_test,Klanjsek,Ward_ladder}.
Clearly this idea of external control of interaction strength of TLL fits well with mesoscopic physics.
Nevertheless it is not fully emphasized thus far.

One of representative mesoscopic systems yielding the TLL is the quantum wire.
With the aid of the long-range Coulomb interaction, the TLL parameter of the quantum wire is expected to be small.
Such a strongly repulsive TLL, or the 1D Wigner crystal~\cite{ftn1,Schulz_WignerCrystal}, in quantum wires is a highly nontrivial electron state
thanks to the interplay of the long-range interaction and the non-Fermi-liquid nature.

\begin{figure}[b!]
 \centering
 \includegraphics[bb= 0 0 900 300, width=\linewidth]{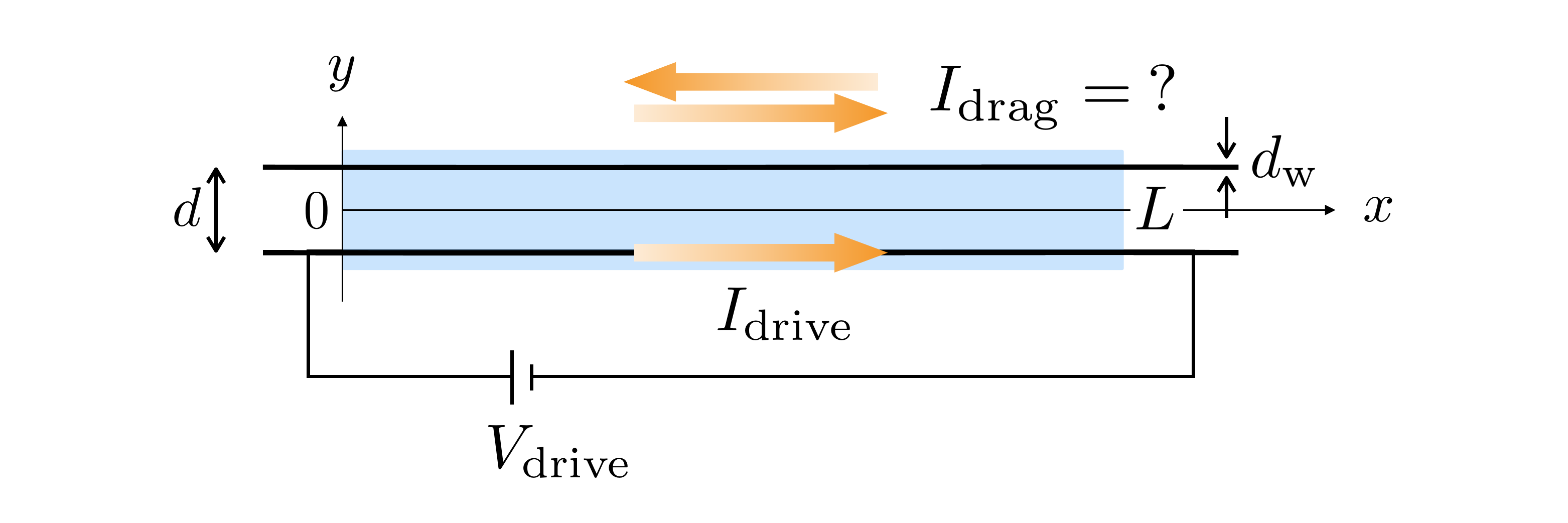}
 \caption{
 Coupled quantum wires (the shaded area).
 }
 \label{wires}
\end{figure}

Recent progresses in fabricating quantum wires have made it possible to
address experimentally a remarkable transport property of TLL, that is, negative Coulomb drag~\cite{Yamamoto, Laroche2011, Laroche2014}.
Coulomb drag in general is an induction phenomenon of the electric current by another capacitively coupled current (Fig.~\ref{wires}),
purely originating from the long-range Coulomb interaction.
Coulomb drag of quantum wires has received intensive theoretical interests~\cite{Flensberg1998,Nazarov1998,Klesse2000,Ponomarenko2000,Pustilnik2003,Pustilnik2006,Peguiron2007} for decades.
These theories predicted the positive drag of parallel drive and drag currents.
However, in sharp contrast to the theoretical predictions,
the negative drag of antiparallel drive and drag currents was observed experimentally~\cite{Yamamoto}.
The negative drag was ``unexpected''~\cite{Yamamoto} in this sense and is still elusive because of lack of theoretical explanation.
The authors of Ref.~\onlinecite{Yamamoto} indicated that the Wigner crystal formed on the drag wire would be responsible for the negative drag.
Later Refs.~\onlinecite{Laroche2011,Laroche2014} reported that an increase of the voltage induces alternately the positive and negative drags.
The Wigner crystallization on the single wire is not likely to explain this re-entrant negative drag~\cite{Laroche2011}. 
Therefore, the mechanism of the negative drag remains unclear despite the existence of the fascinating experimental results.

In this Paper, we propose a simple theory that explains the negative drag clearly.
Its mechanism is summarized as follows.
When the particle density of the drive wire is commensurate with the hole density of the drag wire,
the particle-hole pairing over the wires occurs, leading to \emph{the positive drag of the hole current}, 
that is, the negative drag of the electric current as shown in Fig.~\ref{drag}~(b).
We point out that the strong repulsion of the TLL is necessary to induce the negative drag.
We also clarify the crucial role of the controllability of interactin strength in the negative drag.

Let us consider quantum wires of the length $L$ and the width $\dw$ separated by the distance $d$ (Fig.~\ref{wires}) and
apply the external voltage $V_{\mathrm{drive}}$ to the drive wire in order to induce the drive current $I_{\mathrm{drive}}$.
Our purpose is to see the sign of the ratio $I_{\mathrm{drag}}/I_{\mathrm{drive}}$.
If the interwire interaction was absent, each wire would have a TLL~\footnote{In Eq.~\eqref{H_n}, we dropped the spin degree of freedom because it is irrelevant in our problem as long as we do not apply the magnetic field.}.
\begin{equation}
 \mathcal{H}_n= \frac{\hbar u_n}{2\pi}\int_0^L dx \biggl[ K_n
  (\partial_x\theta_n)^2+\frac{1}{K_n}(\partial_x\phi_n)^2\biggr].
  \label{H_n}
\end{equation}
Here $u_n$ is the velocity of the TLL and $K_n$ ($<1$) is the TLL paramter.
We label the drive and drag wires by $n=1$ and $n=2$.
$\phi_n$ and $\theta_n$ satisfy the commutation relation $[\phi_n(x),\partial_x\theta_m(y)]=i\pi\delta_{n,m}\delta(x-y)$, 
leading to the equation of motion $\partial_t\phi_n=u_nK_n\partial_x\phi_n$.
The Hamiltonian \eqref{H_n} describes the bosonic excitation with the linear dispersion, that is, the TLL.
Since the TLL describes the particle-hole excitation,
the bosonic field $\phi_n$ is related to the fluctuations of the electron density $\rho_n(x)$ from its average $\bar\rho_n$
in the $n$th wire ~\cite{Haldane_bosonization},
\begin{align}
 \rho_n(x) &= \bar\rho_n  -\frac{\partial_x\phi_n}{\pi}
 + 2\bar\rho_n\sum_{p=1}^\infty \cos(2p\pi\bar\rho_nx+2p\phi_n).
 \label{rho}
\end{align}

The interwire interaction induces two effects: the renormalization of the TLL parameters $K_n$ and a locking of bosonic fields over the wires.
The latter needs a careful consideration.
Let us consider the interwire Coulomb interaction $\int drdxV_\perp(r)\rho_1(x)\rho_2(r+x)$.
Although the Coulomb interaction is originally long-ranged, 
it is replaceable to the effective short-range interaction $g\rho_1(x)\rho_2(x)$ in the low-energy limit.
The long-rang nature of the interaction renormalizes the strength $g$ of the effective interaction.
We will come back to this point later.
The short-range intearction $g\rho_1(x)\rho_2(x)$ generates an interaction
$4g\bar\rho_1\bar\rho_2\cos(2p_1\pi\bar\rho_1x+2p_1\phi_1)\cos(2p_1\pi\bar\rho_2x+2p_2\phi_2)$, that is,
\begin{equation}
 \sum_{\nu=\pm}2g\bar\rho_1\bar\rho_2\cos[2\pi(p_1\bar\rho_1+\nu p_2\bar\rho_2)x+2(p_1\phi_1+\nu p_2\phi_2)],
  \label{coscos}
\end{equation}
where $p_1$ and $p_2$ are integers.
In the case of a single quantum wire, these cosine interactions are irrelevant except for 
the Mott insulating case with repulsive interaction, when $\bar\rho=1/a_0$,
where $a_0$ is the lattice spacing of the wires~\cite{Giamarchi_book}.
However, in our model of coupled quantum wires,  $\bar\rho_1$ and $\bar\rho_2$ are independently tunable parameters.
This is important for the negative drag as we discuss below.
In the following we consider the case with $p_1=p_2=p$.
It is easy to extend the results for general cases with $p_1\not=p_2$.

Including these effects, we consider a model with a Hamiltonian
\begin{align}
 \mathcal H
 &= \sum_{\nu=\pm}\biggl[\frac{\hbar u_\nu}{2\pi}\int_0^Ldx\biggl[K_n(\partial_x\theta_\nu)^2
 +\frac{1}{K_\nu}(\partial_x\phi_\nu)^2\biggr]  +V_\nu^p\biggr],
 \label{H}
\end{align}
where $\phi_\pm=(\phi_1\pm\phi_2)/\sqrt{2}$, $\theta_\pm=(\theta_1\pm\theta_2)/\sqrt{2}$,
\begin{equation}
 V_\nu^p = 2g\bar\rho_1\bar\rho_2\int_0^Ldx\cos(4p\pi\bar\rho_\nu x+p\sqrt{8}\phi_\nu), \quad (p\in\mathbb{N}),
  \label{V_pm}
\end{equation}
and $\bar\rho_\pm=(\bar\rho_1\pm\bar\rho_2)/2$.

When $\bar\rho_\nu$ is commensurate, that is, 
\begin{equation}
 2pa_0\bar\rho_\nu\in \mathbb Z,
  \label{comm_general}
\end{equation}
the interaction \eqref{V_pm} survives the spatial integration.
Otherwise the interaction \eqref{V_pm} is negligible because of the destructive oscillation $e^{i2p\bar\rho_\nu x}$.
In the commensurate case \eqref{comm_general}, 
the cosine interaction \eqref{V_pm} causes an energy gap for $\phi_\nu$ and
fixes $\phi_\nu$ to a constant value.
This is called the locking effect.

Electric currents of the symmetric and antisymmetric sectors are given by $J_\pm=-e\langle\partial_t\phi_\pm\rangle/\pi$, 
where $-e<0$ is the electron charge.
Similarly one can define the current $J_n=-e\langle\partial_t\phi_n\rangle/\pi$ on the $n$th wire.
As well as the incommensurate density, the incommensurate current conflicts with the locking effect 
because the current adds a temporal oscillation $e^{iJ_\nu\pi t/e}$ to $V_\nu^p$.
Thus, if $V_\nu^p$ eventually locks the $\phi_\nu$ field over the wires to a constant, then $J_\nu=0$ follows~\cite{suppl}.
Note that two currents $J_\pm$ do not vanish simultaneously.
If $J_+=J_-=0$, the circuit has no current,
which is different from the situation that we consider here.
The drag and drive currents are given by
\begin{equation}
 I_{\mathrm{drag}}= \frac{J_+-J_-}{\sqrt{2}}, \quad I_{\mathrm{drive}} = \frac{J_++J_-}{\sqrt 2}.
  \label{I_drag}
\end{equation}
$J_+=0$ leads to the negative drag $I_{\mathrm{drag}}/I_{\mathrm{drive}}=-1$ 
and $J_-=0$ leads to the positive drag $I_{\mathrm{drag}}/I_{\mathrm{drive}}=1$. 
This is the essence for the negative drag.
If we take $p_1\not=p_2$ in Eq.~\eqref{coscos}, 
the locking of $\phi_\pm\propto(\phi_1\pm(p_2/p_1)\phi_2)$ leads to $I_{\mathrm{drag}}/I_{\mathrm{drive}}=\pm p_2/p_1$.

\begin{figure}[b!]
 \centering
 \includegraphics[bb= 0 0 600 300, width=\linewidth]{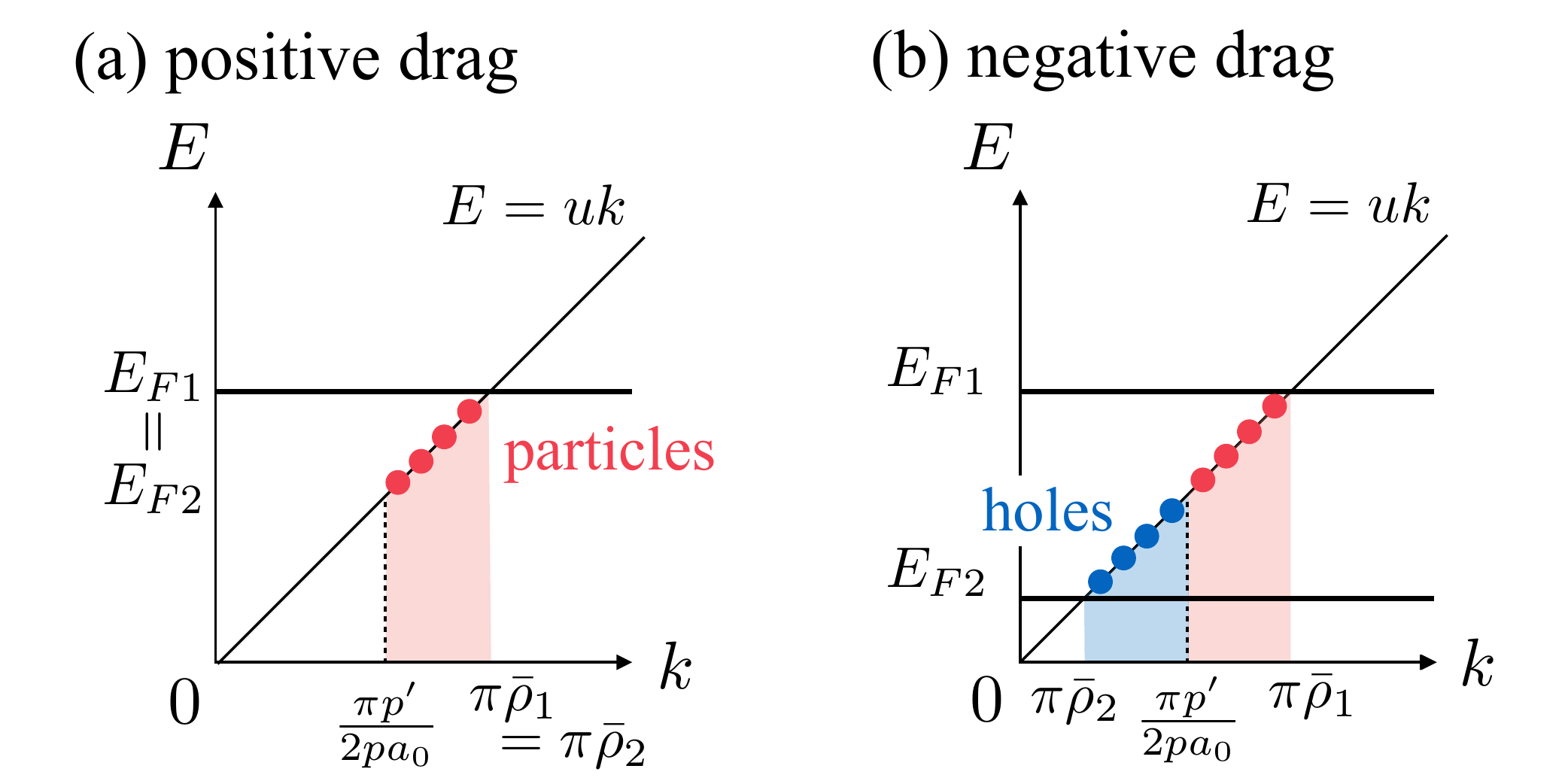}
 \caption{(a) The situation \eqref{p-p} when the positive drag
 occurs. 
 $E_{Fn}$ is the Fermi level of the $n$th wire.
 The drive and drag wires have the equal particle densities.
 (b) The situation \eqref{p-h} when the negative drag occurs.
 The density of doped particles in the drive wire is equal to that 
 of doped holes in the drag wire.
 }
 \label{drag_linear}
\end{figure}

Physical picture becomes clear when we take the particle-hole description.
We assume $\bar\rho_1>\bar\rho_2$ and
\begin{equation}
 \bar\rho_+ = \frac{p'}{2pa_0}.
  \label{commensurate}
\end{equation}
$p$ and $p'\in\mathbb{N}$ are coprime.
Compared to the average filling \eqref{commensurate}, the drive (drag) wire has more (less) electrons.
Thus one may regard the drive and drag wires as particle-doped and hole-doped conductors respectively.
In this particle-hole description,  the condition
\eqref{comm_general} for $\nu=+$ becomes
\begin{equation}
 \rho_1^{\mathrm{p}} = \rho_2^{\mathrm{h}},
  \label{p-h}
\end{equation}
where $\rho_n^{\mathrm{p}}=\bar\rho_n-p'/2pa_0$ and
$\rho_n^{\mathrm{h}}=-\rho_n^{\mathrm{p}}$ are densities 
of doped particles and holes in the $n$th wire.
Thus one can see that the negative drag occurs when the density of doped particles and holes are balanced.
Similarly the positive drag originates from a commensurate condition $\bar\rho_-=0$, that is,
\begin{equation}
 \rho_1^{\mathrm{p}}=\rho_2^{\mathrm{p}}.
  \label{p-p}
\end{equation}
We visualized the conditions \eqref{p-h} and \eqref{p-p} in Fig.~\ref{drag_linear}.
When the condition \eqref{p-p} holds,
the strong interwire interaction locks relative positions of particles over the wires to avoid costing the Coulomb potential energy $V_-^p$, 
resulting in the positive drag [Fig.~\ref{drag}~(a)]. 
On the other hand, when the condition \eqref{p-h} holds,
the strong attraction works between particles on one wire and holes on the other wire, forming particle-hole pairs over the wires.
This particle-hole pairing allows for the positive drag of the hole current, that is, the negative drag [Fig.~\ref{drag}~(b)].

\begin{figure}[b!]
 \centering
 \includegraphics[bb=0 0 1350 450, width=\linewidth]{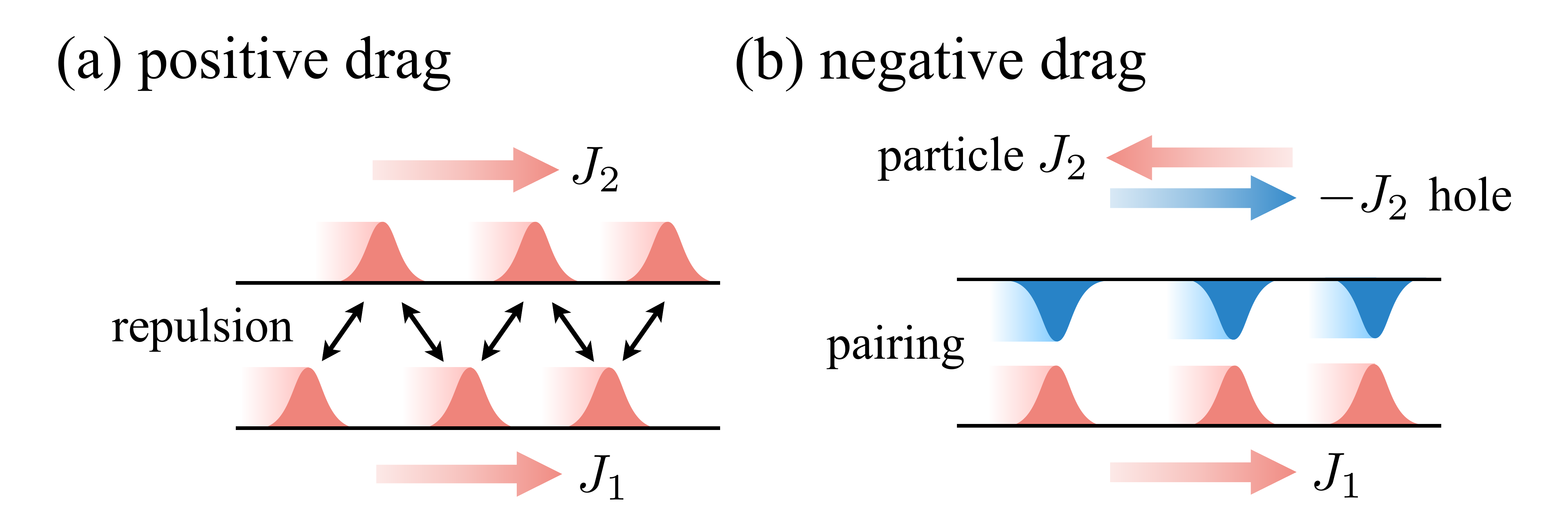}
 \caption{
 Cartoons of (a) the positive and (b) the
 negative drags.
 (a) The interwire repulsion of particles results in the
 global shift of the particles, that is, the positive  drag.
 (b) The negative drag is the positive drag of holes due to the
 particle-hole pairing.
 }
 \label{drag}
\end{figure}

As shown above, the simple model \eqref{H} explains the positive and negative drags on equal footing.
The originality of the present model is in the inclusion of the cosine interaction $V_+^p$ of the symmetric sector.
Except this point, the model \eqref{H} is identical to the models commonly employed in theoretical studies of Coulomb drag~\cite{Nazarov1998,Klesse2000,Ponomarenko2000,Peguiron2007}.
The preceding theories implicitly assumed the incommensurate $\bar\rho_+$
and ignored $V_+^p$~\cite{Flensberg1998,Nazarov1998,Klesse2000,Ponomarenko2000,Pustilnik2003,Pustilnik2006,Peguiron2007}.   
This assumption is certainly natural for 1D conductors, but not always true.
There is no principal reason to rule out the possibility \eqref{commensurate}.
In the following we discuss the situation where the condition \eqref{commensurate} holds.

Since quantum wires are usually far below the unity filling (i.e. $\bar\rho_n\ll1/a_0$),
large $p$ is requisite for satisfying Eq.~\eqref{commensurate}.
In this case, the relevance of the cosine interaction is reduced.
This is because the cosine interaction $V_+^p$ has a scaling dimension $d=2p^2K_+$, and $d<2$ is required in order to lock $\phi_+$.
Larger $p$ makes it more difficult to achieve $d<2$.
In contrast the positive drag does not require the large $p$ because 
the condition \eqref{comm_general} is easily satisfied for $\bar\rho_1=\bar\rho_2$.
This point makes the positive and negative drags unequal.
The strong repulsion $K_+<1/p^2\ll1$ is necessary for the negative drag.

Let us show that the long-range intrawire and interwire Coulomb interactions are the keys for realization of the condition
$K_+<1/p^2\ll1$.
The intrawire Coulomb interaction $\int dxdrV_\parallel(r)\rho_n(x)\rho_n(x+r)$ renormalizes $u_n$ and $K_n$ as follows.
It generates the kinetic term 
$V_\parallel(q=0)(\partial_x\phi_n)^2/\pi^2$,
where $V_\parallel(q)$  is the Fourier transform of $V_\parallel(r)$.
When the electron density is low enough, the Coulomb interaction is unscreened and divergent at $q\to0$.
Such a divergence at long distance is crucial to the drag current flowing through the whole wire.
In order to include the effect of the cut-off of the divergence due to the finite length of the wires,
we model the intrawire interaction as
$V_\parallel(r)=e^2/4\pi\varepsilon\sqrt{r^2+\dw^2}$~\cite{Gold},
where $\varepsilon$ is the dielectric constant of the wires.
Because of the dimensionality, $V_\parallel(q)\simeq(e^2/2\pi\varepsilon)\log(1/|q|\dw)$
exhibits the logarithmic divergence~\cite{Gold,Schulz_WignerCrystal}.
The finite length $L$ cuts off the divergence as $V_\parallel(q\to0)\simeq(e^2/2\pi\varepsilon)\log(L/\dw)$, leading to~\cite{suppl}
\begin{equation}
 K_n\simeq K\equiv\biggl[\frac{e^2}{\pi^2\hbar\varepsilon v_F}
  \log(L/\dw)\biggr]^{-1/2},
  \label{K_n}
\end{equation}
and $u_n=u\equiv v_F/K$.
Here $v_F$ is the Fermi velocity of the free electron.
Since the Coulomb interaction in 1D diverges at both long and short distances, the TLL parameter \eqref{K_n} depends on both $L$ and $\dw$.
The relation \eqref{K_n} shows that we can control the strength of repulsion of the TLL by changing  geometrical parameters $L$ and $\dw$
of the circuit.
One can find a similar argument in Refs.~\onlinecite{Glazman_G_Wigner,Matveev_G_Wigner}.

The interwire interaction is similarly given by $V_\perp(q\to0)\simeq(e^2/2\pi\varepsilon')\log(L/d)$,
where $\varepsilon'$ is the dielectric constant of an insulating medium in between the wires. 
The effective $q$ independence of $V_\perp(q\to0)$ allows us to replace it to an effective short-range interaction 
$V_\perp(r)\simeq V_\perp(q\to0)\delta(r)$ in the low-energy limit $q\to0$, which gives
the strength $g=V_\perp(q\to0)$ of the cosine interaction \eqref{V_pm}.

Thus both of the intrawire and the interwire interactions renormalize the TLL parameter as~\cite{suppl}
\begin{equation}
 K_\pm = \biggl[\frac{e^2}{\pi^2\hbar \varepsilon v_F} \biggl\{\log(L/\dw)\pm
  \frac{\varepsilon}{2\varepsilon'}\log(L/d)\biggr\}\biggr]^{-1/2},
  \label{K_pm}
\end{equation}
and the velocity to $u_\pm=v_F/K_\pm$.
In general $K_+<K_-$ holds.
The TLL parameters \eqref{K_pm} are controllable with the geometrical parameters of the circuit.
When the wires are long ($L\gg\dw$) and separated by a medium with $\varepsilon'\gg\varepsilon$,
The TLL parameter $K_+\sim[\log(L/\dw)]^{-1/2}$ is basically determined only from the intrawire Coulomb intearction.
Since $K_+\sim[\log(L/\dw)]^{-1/2}$ can become arbitrally small,
the condition $K_+<1/p^2\ll1$ for the negative drag can be satisfied with the aid of the long-range 
intrawire Coulomb interaction.

Let us discuss major factors that disturb the negative drag.
They are the temperature and the incommensurability.
First we estimate the temperature effect.
Under the condition \eqref{commensurate}, the symmetric mode acquires an energy gap for $\phi_+$.
We can easily see that an expansion $\cos(p\sqrt{8}\phi_+)\simeq1-4p^2(\phi_+-Q_+^0)^2$ around a bottom $\phi_+=Q_+^0$ of the cosine $V_+^p$
generates the quadratic mass term~\cite{suppl}.
However the locking is weakened by instantons that represents the tunneling of neighboring locking values of 
$Q_+^0\to Q_+^0\pm\pi/p\sqrt{2}$.
Therefore, in order to have Coulomb drag, the excitation of instantons should be suppressed~\cite{Nazarov1998,Ponomarenko2000}.
Fortunately the exact excitation spectrum of the instanton is available~\cite{Dashen,Zamolodchikov},
which is composed of a soliton and an antisoliton.
Their excitation gap, $M$, is exactly derived~\cite{Lukyanov,suppl}.
We can suppress thermal excitations of the instanton at
\begin{equation}
 k_BT\ll M\simeq 
    \frac{2e^2\bar\rho_+}{p\pi^2\varepsilon}\biggl[\frac{2\varepsilon}{\varepsilon'}\log(L/\dw)\log(L/d)\biggr]^{1/2}.
  \label{no_soliton}
\end{equation}
Then $\phi_+$ is well locked to the bottom $Q_+^0$.
As we discussed, the negative drag occurs when $\log(L/\dw)\gg1$ and $\varepsilon/\varepsilon'\ll1$ are satisfied.
Given $\log(L/\dw)\gg\varepsilon'/\varepsilon\gg1$,
the gap $M$ becomes large and then the temperature range \eqref{no_soliton} is wide enough for experimental realizations.

Next we estimate robustness of the negative drag against the incommensurability.
Let us displace $\bar\rho_+$ from the commensurate value \eqref{commensurate}.
We use $\xi_\rho^{-1}=4p\pi|\bar\rho_+-\bar\rho_+^0|$ as a measure of the displacement.
Nonzero $\xi_\rho^{-1}$ adds the incommensurate oscillation $e^{ix/\xi_\rho}$ to the cosine interaction \eqref{V_pm},
which disturbs the coherence of $Q_+^0$.
Let us write the lowest-energy excitation gap of $\phi_+$ as $u_+/\xi$ ($\not= M$)~\cite{suppl}.
$\xi$ gives the coherent interval of the locking of $\phi_+$.
When $\xi<\xi_\rho$, the oscillation is very slow over the length $\xi$ , and the locking by $V_+^p$ persists.
On the other hand, when $\xi>\xi_\rho$, the oscillation dissolves the locking.
Therefore the negative drag lasts for $\xi_\rho^{-1}<\xi^{-1}$, that is~\cite{suppl},
\begin{equation}
 \frac{|\bar\rho_+-\bar\rho_+^0|}{\bar\rho_+}
 <\mathcal{A}\equiv\pi\biggl[\frac{2\varepsilon}{\varepsilon'}\frac{\log(L/d)}{\log(L/\dw)}\biggr]^{1/2}
  \label{cond_negative}
\end{equation}
for $\epsilon'/\epsilon\gg1$.

Furthermore, in order to realize the negative drag, the positive drag should be suppressed.
Since the positive drag is stabilized for
\begin{equation}
  \frac{|\bar\rho_1-\bar\rho_2|}{\bar\rho_+}<\mathcal{A},
   \label{cond_positive}
\end{equation}
for $\epsilon'/\epsilon\gg1$,
the electron densities must break the inequality \eqref{cond_positive}.
Thus the negative drag occurs when $\bar\rho_1$ and $\bar\rho_2$ satisfy
$|\bar\rho_+-\bar\rho_+^0|<\mathcal{A}\bar\rho_+<|\bar\rho_1-\bar\rho_2|$.

Let us compare our theory with the existing experiments.
According to \eqref{K_n} and \eqref{K_pm}, we can prepare $K_\pm$ as small as we wish by taking large enough $L/\dw$.
In fact an experiment shows $K_-\simeq0.08\pm0.02$~\cite{Laroche2014}.
Such a small TLL parameter is impossible without the long-range repulsion.
For instance, the 1D Hubbard model only with the on-site repulsion  has the TLL parameter larger than $1/2$
for any filling and parameters~\cite{Schulz_Hubbard}. 
The extraordinary small TLL parameter observed experimentally indicates that
the long-range nature of the Coulomb interaction does exist in quantum wires.

In general the $N$-component TLL on the quantum wire has a quantized conductance $G=Ne^2/h$~\cite{Tarucha_G}.
When the spin is included, $G=N'(2e^2/h)$ with $N=2N'$.
In our model, the interaction $V_\nu^p$ locks a half degree of freedom in $\phi_2=(\phi_+-\phi_-)/\sqrt{2}$,
resulting in a fractionalization $G=(1/2)e^2/h$, similarly to Ref.~\onlinecite{Pustilnik2006}.
Inclusion of the spin degree of freedom doubles it to $G=e^2/h$, which is basically consistent with experiments~\cite{Laroche2014}.
However the height is not well quantized as $e^2/h$ for some situations:
Fig.~3a of Ref.~\onlinecite{Laroche2011} shows that the height of conductance plateaus are reduced.
Such reduction is attributed to, for example, the impurity~\cite{Ogata_dirty_TLL} and the spin degree of freedom~\cite{Matveev_G_Wigner}
and irrelevant in the emergence of the negative drag.
Our theory explains at least the negative drag on the lowest conductance plateau observed in Fig.~4a of Ref.~\onlinecite{Laroche2011}.
Those on higher plateaus can be explained after a straightforward extension.
When intrawire intercomponent interactions are negligible, which is usually true in the low-energy limit, 
we can easily extend our model to the $N$-component TLL on higher conductance plateaus.

Figure~\ref{drag_linear} implies that the positive and negative drag occurs alternately as $\bar\rho_1$ and $\bar\rho_2$ are increased 
independently.
Besides, Coulomb drag becomes less prominent in larger $\bar\rho_+$.
Large enough $\bar\rho_+$ easily satisfies $\max\{|\bar\rho_+-\bar\rho_+^0|, |\bar\rho_1-\bar\rho_2|\}<\mathcal{A}\bar\rho_+$,
inducing the positive and negative drags simultaneously [Eqs.~\eqref{cond_negative} and \eqref{cond_positive}], that is, no drag as a whole.
Such qualitative dependences on the densities $\bar\rho_1$ and $\bar\rho_2$ are consistent with
the experiments~\cite{Laroche2011,Laroche2014} by translating gate voltages to electron densities.

We are grateful to C. Berthod, T. Giamarchi and E. Iyoda for
stimulating discussions. 
S.C.F. was supported by the Swiss SNF under Division II.

\newpage

\begin{table}[t!]
{\bf \large Supplemental Material for \\
``Negative Coulomb Drag in Coupled Quantum Wires''}
\end{table}

\section{Locking of the zero mode}

Here we show that the locking of $\phi_+$ leads to $\langle
\partial_t\phi_+\rangle=0$.
To see this, we introduce a mode expansion~\cite{Giamarchi_book},
\begin{align}
 \phi_+(x)
 &= Q_+ + \frac{\pi N_+}L x 
 \notag \\
 & \qquad
  + \sum_{q\not=0} \sqrt{\frac{\pi K_+}{2L|q|}} \,
 \bigl(a_{q,+}e^{iqx}+ a_{q,+}^\dagger e^{-iqx}\bigr),
 \label{mode_phi} \\
 \theta_+(x)
 &= \Theta_+ + \frac{\pi P_+ }L x 
 \notag \\
 & \qquad
+ \sum_{q\not=0} \operatorname{sgn}(-q) \sqrt{\frac{\pi
 K_+^{-1}}{2L|q|}}
 \, \bigl(q_{q,+}e^{iqx} + a_{q,+}^\dagger e^{-iqx}\bigr).
 \label{mode_theta}
\end{align}
The first lines of Eqs.~\eqref{mode_phi} and \eqref{mode_theta}
represent the zero modes of $\phi_+(x)$ and $\theta_+(x)$, where 
$(Q_+, \Theta_+)$ and $(P_+, N_+)$ are canonical conjugate, that is,
$[Q_+, P_+]=[\Theta_+, N_+] = i$.
The second lines represent the $q\not=0$ modes.
In the absence of the cosine interaction, $a_{q,+}$ is an annihilation
operator of the TLL.
Under the condition (6) of the main text, the effective Hamiltonian
of the symmetric mode becomes
\begin{align}
 \mathcal{H}_+
 &= \frac {\hbar u_+}{2\pi}\int dx \biggl[ K_+(\partial_x\theta_+)^2 
  +\frac 1{K_+}(\partial_x\phi_+)^2\biggr] \notag \\
 & \qquad + 2g\bar\rho_1\bar\rho_2 \int dx\cos(p\sqrt{8}\phi_+)
 \label{H_+}
\end{align}
with
\begin{equation}
 g = V_\perp(q\to0) = \frac{e^2}{2\pi\varepsilon'}\log(L/d).
\end{equation}
The mode expansions \eqref{mode_phi} and \eqref{mode_theta} lead to
\begin{align}
 \mathcal H_+
 &= \frac{\pi\hbar u_+}{2K_+L} N_+^2 + \frac{\pi\hbar u_+K_+}{2L} P_+^2 
 + \sum_{q\not=0} \hbar u_+|q| a_{q,+}^\dagger a_{q,+}.
\end{align}
In the presence of the cosine interaction, the TLL acquires the
excitation gap.
When the temperature is much lower than the energy gap of the soliton, 
we can expand the cosine as
$\cos(p\sqrt{8}\phi_+)\simeq 1-4p^2(\phi_+-Q_+^0)^2$ with a constant $Q_+^0$. 
Then the Hamiltonian \eqref{H_+} turns into
\begin{align}
 \mathcal H_+
 &=\frac{\pi\hbar u_+}{2K_+L} N_+^2+
 \frac{\pi\hbar u_+K_+}{2L} P_+^2 +\frac{LM_1^2}{2\pi\hbar u_+K_+}(Q_+-Q_+^0)^2
 \notag \\
 & \qquad + \sum_{q\not=0} \hbar\sqrt{(u_+q)^2+(M_1/\hbar)^2} b_{q,+}^\dagger
 b_{q,+}.
 \label{H_+_gapped}
\end{align}
We introduced the creation and annihilation operators, $b_{q,+}^\dagger$
and $b_{q,+}$, through a Bogolioubov transformation,
\begin{equation}
 \begin{pmatrix}
  b_{q,+} \\
  b_{-q,-}^\dagger
 \end{pmatrix}
 =
 \begin{pmatrix}
  \cosh\theta_{q,+} & -\sinh\theta_{q,+} \\
  -\sinh\theta_{q,-} & \cos\theta_{q,-}
 \end{pmatrix}
 \begin{pmatrix}
  a_{q,+} \\
  a_{-q,+}^\dagger
 \end{pmatrix},
\end{equation}
with $ \coth(2\theta_{q,-})\simeq1+2({\hbar u_+q}/{M_1})^2$.
Since there is no Bose-Einstein condensate, that is, $\langle
b_{q,+}\rangle=\langle b_{q,+}^\dagger\rangle =0$, the average $\langle
\partial_t\phi_+\rangle= u_+K_+\langle \partial_x\theta_+\rangle$ is
fully determined from the zero mode,
\begin{equation}
 \langle \partial_t\phi_+\rangle
  = \frac{\pi u_+K_+}L \langle P_+\rangle.
\end{equation}
As one can see from Eq.~\eqref{H_+_gapped}, the zero mode $P_+$ is
subject to the Hamiltonian of a harmonic oscillator,
\begin{equation}
 \mathcal H_+^{\rm zero} 
  =\frac{\pi\hbar u_+}{2K_+L}N_+^2
  +\frac{\pi\hbar u_+K_+}{2L} P_+^2 +\frac{LM_1^2}{2\pi\hbar u_+K_+}(Q_+-Q_+^0)^2.
  \label{harmonic}
\end{equation}
Note that the zero mode $N_+$ is independent of both $P_+$ and $Q_+$.
Although Eq.~\eqref{harmonic} is, of course, the trivial consequence of
the quadratic approximation of the cosine, it leads to an important
results,
\begin{equation}
 \langle P_+ \rangle = \langle (Q_+-Q_+^0)\rangle=0.
  \label{zero_average}
\end{equation}
$\langle \partial_t\phi_+\rangle = 0$ immediately follows from
Eq.~\eqref{zero_average}. 
The relation \eqref{zero_average} holds only when the quadratic
approximation of the cosine is valid.

\section{Gap of the soliton and the antisoliton}

The field theory with the Hamiltonian \eqref{H_+} is called as the
sine-Gordon theory.
Since the sine-Gordon theory is integrable, its various quantities are
exactly available.
The excitation gap of the soliton and the antisoliton at $q=0$ is
given by~\cite{Lukyanov}
\begin{equation}
 M=\frac{2\hbar u_+\Gamma(\frac{\gamma}2)}{a_0\sqrt{\pi}\Gamma(\frac{1+\gamma}2)}\biggl(\frac{a_0^2\pi g\bar\rho_1\bar\rho_2\Gamma(\frac1{1+\gamma})}{\hbar u_+\Gamma(\frac{\gamma}{1+\gamma})}\biggr)^{(1+\gamma)/2}.
\end{equation}
For $p^2K_+\ll1$, the parameter $\gamma=1/[(p^2K_+)^{-1}-1]\simeq
p^2K_+$ is also small.
Taking the lowest order of $\gamma\simeq p^2K_+$, we obtain
\begin{equation}
 M\simeq \frac 4{p}\biggl(\frac{\hbar u_+g\bar\rho_1\bar\rho_2}{\pi K_+}\biggr)^{1/2},
\end{equation}
where we used the features, $\Gamma(1)=1$ and
$\Gamma(z)\Gamma(1-z)=\pi/\sin(\pi z)$, of the Gamma function $\Gamma(z)$.
$u_\pm/K_\pm$ and $u_\pm K_\pm$ satisfy
\begin{align}
 \frac{u_\pm}{K_\pm}
 &= \frac uK\pm\frac{1}{\pi\hbar} V_\perp(q=0) \notag \\
 &= v_F+\frac{2}{\pi\hbar} V_\parallel(q=0)
 \pm\frac{1}{\pi\hbar} V_\perp(q=0) \notag \\
 &\simeq \frac{e^2}{\pi^2\hbar\varepsilon} \log(L/\dw) \pm \frac{e^2}{2\pi^2\hbar\varepsilon'}\log(L/d)
\end{align}
and $u_\pm K_\pm=uK=v_F$, resulting in
\begin{align}
 K&=\biggl[\frac{e^2}{\pi^2\hbar\varepsilon v_F}\log(L/\dw)\biggr]^{-1/2}, \\
 K_\pm &= K\biggl[1\pm \frac{\varepsilon}{2\varepsilon'}\frac{\log(L/d)}{\log(L/\dw)}\biggr]^{-1/2}
\end{align}
and $u_\pm=v_F/K_\pm$.
In the end, when $\varepsilon'/\varepsilon\gg1$, the gap of the soliton is approximated to
\begin{equation}
 M\simeq 
    \frac{2e^2\bar\rho_+}{p\pi^2\varepsilon}\biggl[\frac{2\varepsilon}{\varepsilon'}\log(L/\dw)\log(L/d)\biggr]^{1/2}.
\end{equation}
Then the lowest-energy excitation gap $M_1$ is reduced to
\begin{align}
 M_1
 &=2M\sin(\pi\gamma/2) \notag \\
 &\simeq 
 2p\bar\rho_+\biggl[\frac{2\pi^2e^2\hbar v_F}{\varepsilon'}\log(L/d)\biggr]^{1/2}.
 \label{M_1}
\end{align}
Now it is straightforward to derive
\begin{equation}
 \frac{M_1}{2\pi\hbar pu_\pm}\simeq
 \pi\bar\rho_+\biggl[\frac{2\varepsilon}{\varepsilon'}\frac{\log(L/d)}{\log(L/\dw)}\biggr]^{1/2}.
\end{equation}
The condition $\xi_\rho^{-1}<\xi^{-1}$ in the main text can be rewritten as
\begin{equation}
 \biggl|\bar\rho_1+\bar\rho_2-\frac{p'}{pa_0}\biggr| < \frac{M_1}{2\pi\hbar u_+},
\end{equation}
which becomes Eq.~(14) in the main text.
On the other hand, the corresponding condition for the positive drag is
\begin{equation}
 |\bar\rho_1-\bar\rho_2|<\frac{M_1}{2\pi\hbar u_-}\simeq \pi\bar\rho_+\biggl[\frac{2\varepsilon}{\varepsilon'}\frac{\log(L/d)}{\log(L/\dw)}\biggr]^{1/2}.
\end{equation}


\begin{thebibliography}{31}
\bibitem [{ftn()}]{ftn1}%
  \BibitemOpen
  \href@noop {} {}\bibinfo {note} {Note that the 1D Wigner crystal is not the
  true crystalline state because of the absence of the long-range charge order.
  In this Paper we use the general terminology of the strongly repulsive TLL
  for the 1D Wigner crystal in order not to confuse it with the Wigner crystal
  in higher dimensions}\BibitemShut {NoStop}%
\end{thebibliography}
\end{document}